\documentstyle[12pt]{article}
\makeatletter
\def\bbox#1{%
\leavevmode\text{%
\textfont0 \the\textfont\bffam
\scriptfont0 \the\scriptfont\bffam
\scriptscriptfont0 \the\scriptscriptfont\bffam
\@temptokena\everymath \boldmath \everymath\@temptokena
{$\m@th\relax#1$}%
}%
}
\makeatother
\begin{document}

\begin{center}
July 1994\hfill Pisa preprint IFUP-TH 44/94
\end{center}
\begin{center}
\Large Comment on A.~Patrascioiu and E.~Seiler's paper\\
{\bf ``Nonuniformity of the $\bbox{1/N}$ Expansion\\
for Two-Dimensional $\bbox{{\bf O}(N)}$ Models''}.

\bigskip

Massimo Campostrini and Paolo Rossi, \\
\large\it Dipartimento di Fisica dell'Universit\`a and I.N.F.N.,\\
I-56126 Pisa, Italy.
\end{center}

\bigskip

In the persistent absence of a rigorous lattice construction of
two-dimen\-sional asymptotically-free ${\rm O}(N)$-symmetric models
for fixed finite values of $N$, the convergence properties of such
approximation methods as perturbation theory and the $1/N$ expansion
are certainly worth a deep and open-minded investigation.  In this
spirit we shall try to further clarify and qualify some results of
ours, that have been in our opinion misrepresented in Patrascioiu and
Seiler's paper \cite{PS-preprint}.

\medskip

In the {\bf continuum}, we have shown \cite{BCR-ON,CR-review} that
${\rm O}(N)$-symmetric two-dimensional models can be regularized and
renormalized to $O(1/N)$.  Physical quantities can be computed {\it
  analytically\/} with the same precision (including energy, magnetic
susceptibility, mass gap and renormalization-group $\beta$-function)
and are found to agree with the {\it exact\/} (albeit non-rigorous)
results found by Bethe-Ansatz methods and with continuum perturbation
theory results.

The lack of uniform asymptoticity is a {\it predicted property\/} of
the models for quantities diverging in the $\beta\to\infty$ limit; it
would be absent in a $1/(N-2)$ expansion, and in any case it does not
affect the adimensional ratios of physical quantities that are the
sole {\it testable\/} predictions in quantum field theories with no
extrinsic mass scale.

In particular, the so-called mass gap -- $\Lambda$-parameter ratio
(where the $\Lambda$ parameter is defined in standard perturbation
theory) has been computed exactly by Hasenfratz {\it et al.}
\cite{H-Lambda-1,H-Lambda-2}
\begin{equation}
{m\over\Lambda_{\overline{\rm MS}}} =
\left(8\over e\right)^{\textstyle{1\over N-2}}\,
{1\over\displaystyle\Gamma\!\left(1+{1\over N-2}\right)}\,.
\label{Lambda}
\end{equation}
This quantity is series-expandable in $1/N$, and the resulting series
has radius of convergence $1/2$.  Furthermore, Eq.~(\ref{Lambda})
agrees with the result of the $1/N$ expansion \cite{BCR-ON}.

\medskip

In {\bf lattice} ${\rm O}(N)$-symmetric models with standard lattice
action, we have shown that computations to $O(1/N)$ can be carried
down to the level of integral representations of the results.  The
scaling properties can be obtained from an asymptotic series expansion
in the powers of $m^2$, and the following conclusions can be reached.
\begin{enumerate}
\item Physical quantities do not show uniform asymptoticity; however
  the coefficients of the linear and logarithmic growth in the inverse
  coupling $\beta$ can be computed {\it analytically\/} (thus
  bypassing the need for numerical evaluation followed by fitting or
  extrapolation), and are found in agreement with their continuum
  counterparts, which in turn are explicitly predicted by asymptotic
  freedom and perturbation theory.
\item The {\it difference\/} between lattice and continuum evaluation
  of physical quantities shows uniform asymptoticity; moreover, its
  integral representation is series-expandable in $1/\beta$ and the
  coefficients are in agreement with lattice perturbation theory
  predictions.
\item Adimensional {\it ratios\/} of physical quantities, computed in
  the lattice model, show uniform asymptoticity (actually complete
  independence of $\beta$ in the scaling limit) and agree with the
  corresponding continuum results.
\end{enumerate}

\medskip

Our results can be generalized to variant lattice actions and to
lattice models with different symmetries and field content.  There is
no evidence whatsoever of a loss of asymptotic freedom or of
contradiction with perturbation theory results as long as the lattice
and continuum models possess the same symmetry group (including
discrete symmetries).  There is no evidence of obstructions to an
extension to higher orders in $1/N$.

To the best of our understanding, our results (supported by all
available numerical evidence) seem to indicate that, for sufficiently
large $N$ (which we {\it believe\/} to be $N>2$, cfr.\
Eq.~(\ref{Lambda})), two-dimensional lattice ${\rm O}(N)$-symmetric
models are connected by a regular renormalization-group trajectory to
their continuum counterparts, which in turn are asymptotically free
theories whose on-shell behavior and thermodynamic properties are
correctly described by the Bethe-Ansatz solution.


\begin{thebibliography}{1}

\bibitem{PS-preprint}
A.~Patrascioiu and E.~Seiler,
{\it Nonuniformity of the $1/N$ Expansion for Two-Dimensional
${\rm O}(N)$ Models}, preprint MPI-PhT/94-44 (June 1994),
hep-lat/9407003.

\bibitem{BCR-ON}
P.~Biscari, M.~Campostrini, P.~Rossi,
Phys.~Lett.~{\bf 242B}~(1990)~225.

\bibitem{CR-review}
M.~Campostrini and P.~Rossi,
Riv.~Nuovo~Cim.~{\bf 16}~(n.~{\bf 6})~(1993)~1.

\bibitem{H-Lambda-1}
P.~Hasenfratz, M.~Maggiore, and F.~Niedermayer,
Phys.~Lett.~{\bf 245B}, 522 (1990).

\bibitem{H-Lambda-2}
P.~Hasenfratz and F.~Niedermayer,
Phys.~Lett. {\bf 245B}, 529 (1990).

\end{thebibliography}
\end{document}